\documentclass[sigconf,natbib=false]{acmart}
\usepackage{bbding}
\usepackage{enumitem}

\AtBeginDocument{%
  }

\copyrightyear{2026}
\acmYear{2026}
\setcopyright{cc}
\setcctype{by}
\acmConference[SIGIR '26]{Proceedings of the 49th International ACM SIGIR Conference on Research and Development in Information Retrieval}{July 20--24, 2026}{Melbourne, VIC, Australia}
\acmBooktitle{Proceedings of the 49th International ACM SIGIR Conference on Research and Development in Information Retrieval (SIGIR '26), July 20--24, 2026, Melbourne, VIC, Australia}
\acmDOI{10.1145/3805712.3808512}
\acmISBN{979-8-4007-2599-9/2026/07}

\RequirePackage[
  datamodel=acmdatamodel,
  style=acmnumeric,
  ]{biblatex}

\begin{document}

\title{RAD-DPO: Robust Adaptive Denoising Direct Preference Optimization for Generative Retrieval in E-commerce}

\author{Zhiguo Chen}
\authornote{Equal contribution.}
\affiliation{%
  \institution{JD.com}
  \city{Beijing}
  \country{China}}
\email{chenzhiguo22@jd.com}

\author{Guohao Sun}
\authornotemark[1]
\affiliation{%
  \institution{Peking University}
  \city{Beijing}
  \country{China}}
\email{sun_gh@stu.pku.edu.cn}

\author{Yiming Qiu}
\authornote{Corresponding author}
\affiliation{%
  \institution{JD.com}
  \city{Beijing}
  \country{China}}
\email{qiuyiming3@jd.com}

\author{Xingzhi Yao}
\affiliation{%
  \institution{JD.com}
  \city{Beijing}
  \country{China}}
\email{yaoxingzhi1@jd.com}
\author{Mingming Li}
\affiliation{%
  \institution{Institute of Information Engineering, Chinese Academy of Sciences}
  \city{Beijing}
  \country{China}}
\email{limingming@iie.ac.cn}

\author{Huimu Wang}
\authornotemark[2]
\affiliation{%
  \institution{JD.com}
  \city{Beijing}
  \country{China}}
\email{wanghuimu1@jd.com}

\author{Yangqi Zhang}
\affiliation{%
  \institution{JD.com}
  \city{Beijing}
  \country{China}}
\email{zhangyangqi1@jd.com}

\author{Songlin Wang}
\affiliation{%
  \institution{JD.com}
  \city{Beijing}
  \country{China}}
\email{wangsonglin3@jd.com}

\author{Sulong Xu}
\affiliation{%
  \institution{JD.com}
  \city{Beijing}
  \country{China}}
\email{xusulong@jd.com}

\renewcommand{\shortauthors}{Zhiguo Chen et al.}

\begin{abstract}

Generative Retrieval (GR) is rapidly transforming e-commerce search by replacing traditional multi-stage pipelines with the autoregressive decoding of structured Semantic IDs (SIDs). Despite this architectural efficiency, aligning GR models with nuanced, real-world user preferences remains a critical challenge. While Direct Preference Optimization (DPO) offers an efficient alignment solution, its direct application to structured SIDs suffers from three limitations: (i) it penalizes shared hierarchical prefixes, causing gradient conflicts; (ii) it is vulnerable to noisy pseudo-negatives from implicit feedback; and (iii) in multi-label queries with multiple relevant items, it exacerbates a probability "squeezing effect" among valid candidates. To address these issues, we propose RAD-DPO, which introduces token-level gradient detachment to protect prefix structures, similarity-based dynamic reward weighting to mitigate label noise, and a multi-label global contrastive objective integrated with global SFT loss to explicitly expand positive coverage. Extensive offline evaluations and large-scale online A/B testing on JD.com's core search engine demonstrate that RAD-DPO achieves significant improvements in both retrieval precision and training efficiency, proving its robustness for massive industrial deployments.

\end{abstract}

\ccsdesc[500]{Information systems}
\ccsdesc[500]{Information systems~Novelty in information retrieval}
\ccsdesc[500]{Information systems~Information retrieval}
\ccsdesc[500]{Information systems~Retrieval models and ranking}

\keywords{Preference alignment, DPO, Generative retrieval}



\maketitle

\section{Introduction}
\begin{figure*}[htbp]  
    \centering  
    \includegraphics[width=\textwidth]{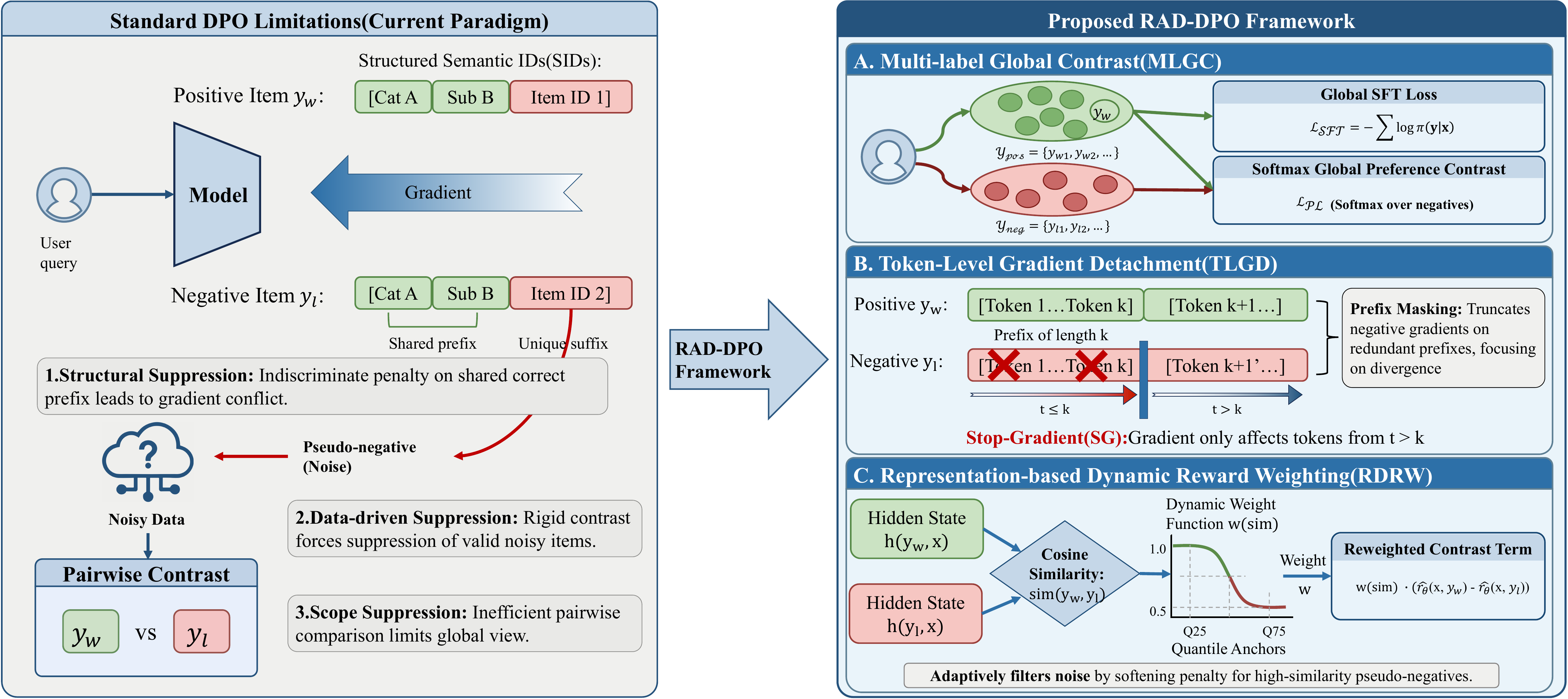}
    \caption{Overview of the RAD-DPO framework. It addresses standard DPO limitations via three core modules: Multi-label Global Contrast(MLGC), Token-Level Gradient Detachment (TLGD) and Representation-based Dynamic Reward Weighting (RDRW).}  
    \label{fig:rad-dpo}  
\end{figure*}


The landscape of Information Retrieval is experiencing a fundamental paradigm shift towards end-to-end Generative Retrieval (GR) \cite{li2024generative, rajput2023recommender, chen2025unisearch, chen2025onesearch, deng2025onerec, zhang2026onemall, zhang2025gpr, fu2025forge, kuai2024breaking}. Unlike traditional multi-stage ranking pipelines, GR simplifies the architecture by formulating item retrieval directly as the autoregressive generation of structured Semantic IDs (SIDs). To represent the vast corpus of items within this generative framework, techniques such as clustering (e.g., DSI \cite{tay2022transformer} and NCI \cite{wang2022neural}) and quantization (e.g., RQ-Kmeans \cite{deng2025onerec} and RQ-VAE \cite{rajput2023recommender}) have proven highly effective in constructing these SIDs. However, while the structural foundation of the ID space is well-established, robustly aligning these GR models with nuanced and dynamic user preferences remains a critical challenge.

To address this alignment bottleneck, Direct Preference Optimization (DPO) has emerged as the prevailing paradigm \cite{rafailov2023dpo, nath2025dpl, wu2024beta, zeng2024token, yang2025token, li2025ambiguity, xiao2024cal, chen2025unisearch, deng2025onerec}. By circumventing the complex reward model training required in traditional reinforcement learning approaches, DPO offers a more direct and stable pathway to optimize generation probabilities. Driven by the demands of large-scale systems, the DPO framework has recently undergone significant evolution. On one hand, researchers have developed reference-free variants \cite{meng2024simpo, hong2024orpo} to substantially improve computational efficiency. On the other hand, list-wise extensions \cite{chen2024softmax, guo2025onesug, chen2025onesearch} have been introduced to accommodate complex multi-negative scenarios, moving beyond simple pairwise comparisons to better capture relative preferences among multiple candidates.

While e-commerce search environments provide abundant implicit preference logs (e.g., user clicks and orders) \cite{lu2025lore, tang2025reaseq, zhang2025pebr, qiu2022pre}, directly applying standard DPO to SIDs introduces a fundamental failure mode: the \textbf{unintended suppression of positive semantic signals} \cite{ren2024learning}. Standard DPO was originally designed for free-text generation, where output tokens are relatively independent. SIDs, by contrast, demand strict adherence to a predefined hierarchical taxonomy. Consequently, DPO's uniform penalty mechanism becomes overly blunt when applied to these structured identifiers, manifesting in three critical optimization bottlenecks:

\begin{itemize}[leftmargin=*, nosep]
    \item \textbf{Gradient Conflict on Shared Prefixes:} Due to the strict hierarchical nature of SIDs, items within the same broad category share identical prefix tokens \cite{rajput2023recommender}. In real-world e-commerce interactions, positive and negative items frequently fall under similar categories; empirically, 34.6\% of our training pairs share the first SID token, and 1.9\% share the first two. Because standard DPO operates at the sequence level, it penalizes the entire negative sequence indiscriminately. This forces the model to suppress shared, correct prefix tokens \cite{rafailov2023dpo}, creating simultaneous "push and pull" gradient signals that cause severe instability during taxonomy generation.
    
    \item \textbf{Vulnerability to Pseudo-Negatives:} Implicit feedback in industrial systems is notoriously noisy. Items that are exposed but unclicked are routinely treated as negative samples, yet this lack of interaction is often due to position bias, visual presentation, or mere click randomness rather than true irrelevance \cite{joachims2017accurately, chen2023bias, zhou2025counterfactual}. Because standard DPO aggressively minimizes the likelihood of all designated negative samples, it heavily penalizes these mislabeled "pseudo-negatives." This rigid over-correction severely distorts the semantic representations of otherwise valid items in the model's latent space.
    
    \item \textbf{Inadequate Multi-Label Coverage:} E-commerce search queries are intrinsically multi-label; a single user prompt typically corresponds to multiple relevant items that could simultaneously satisfy the user's intent. Standard DPO relies on narrow, pairwise comparisons (one positive versus one negative). When this partial-order logic is applied to multi-positive contexts, boosting the probability of one valid candidate inevitably suppresses the probabilities of other concurrent valid candidates. This exacerbates a probability "squeezing effect" \cite{ren2024learning}, restricting the model's global perspective and ultimately degrading overall ranking recall.
\end{itemize}

We propose the Robust Adaptive Denoising Direct Preference Optimization (RAD-DPO) framework (Figure \ref{fig:rad-dpo}) as a systematic solution against these suppression effects. RAD-DPO introduces three targeted modules corresponding to the aforementioned bottlenecks:
First, \textit{Token-Level Gradient Detachment} tackles \textbf{gradient conflicts on shared prefixes} by truncating conflicting backward signals from negative samples, thereby preserving structural integrity. Second, the model's \textbf{vulnerability to noisy pseudo-negatives} is mitigated through \textit{Representation-based Dynamic Reward Weighting}, which adaptively softens optimization penalties guided by semantic similarity. Finally, addressing \textbf{inadequate multi-label coverage}, \textit{Multi-label Global Contrast} replaces myopic pairwise comparisons with a global objective that jointly maximizes the probability mass across all concurrent valid candidates.

In summary, this work makes three primary contributions: (1) we formally identify the unintended suppression of positive signals as a critical failure mode when applying standard DPO to structured generative retrieval; (2) we propose the RAD-DPO framework to systematically mitigate prefix conflicts, data-driven noise, and narrow pairwise scopes; and (3) through extensive offline evaluations and rigorous online A/B testing on JD.com's massive-scale search engine, we demonstrate that RAD-DPO translates theoretical robustness into significant real-world business revenue.

\begin{figure}[t]
    \centering
   \includegraphics[width=\linewidth]{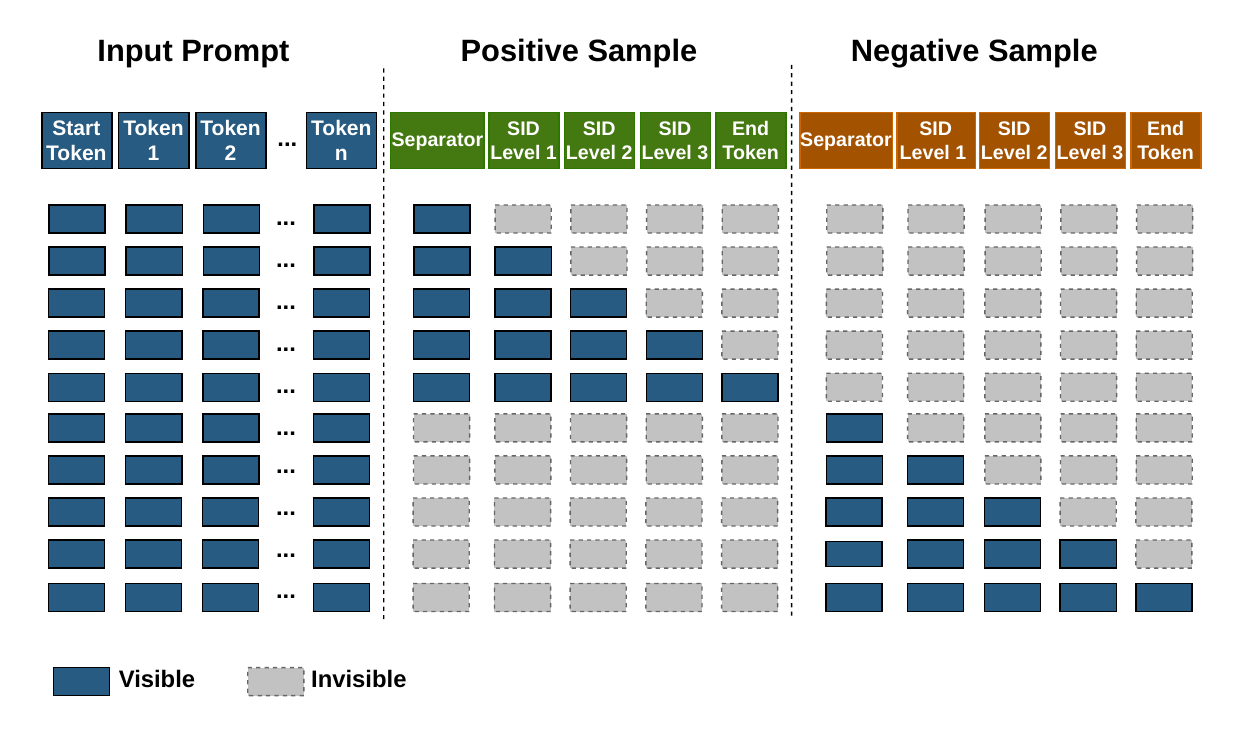}

    \caption{Custom block-diagonal attention mask design for multi-label global contrast.}
    \label{fig-mask}
\end{figure}

\section{Methodology}

In this section, we present the proposed optimization framework designed for structured generation in e-commerce scenarios. The framework consists of three key components: Multi-label Preference Learning with Global Contrast, Token-Level Gradient Detachment, and Representation-based Dynamic Reward Weighting.

\subsection{Multi-label Global Contrast (MLGC)}
To align generative retrieval with the complex multi-interest nature of e-commerce, we construct training samples from search interaction logs. For each session with prompt $x$, items clicked or ordered by the user form the positive set $\mathcal{Y}_{pos} = \{y_{w1}, y_{w2}, \dots\}$. The negative set $\mathcal{Y}_{neg}$ is composed of items that were exposed but not clicked within the same session, augmented by random samples from the global candidate pool to ensure a robust contrastive signal. To balance efficiency and gradient stability, the number of negative samples is fixed to $n$ ($n=3$ in our experiments). 

Traditional Pairwise DPO is restricted to handling single $(x, y_w, y_l)$ pairs, which leads to significant computational redundancy as the prompt $x$ and positive sample $y_w$ are repeatedly re-encoded for every negative candidate. Moreover, in multi-label scenarios, pairwise contrast can exacerbate a probability "squeezing effect" \cite{ren2024learning} among valid candidates. To address this, we design a hybrid multi-label contrastive loss: 
\begin{equation}
\mathcal{L}_{total} = \mathcal{L}_{SFT} + \mathcal{L}_{PL}
\end{equation}

\paragraph{Multi-label SFT Enhancement.} To fully leverage information from multiple positive samples, we introduce a global SFT loss within the positive data cluster:
\begin{equation}
\mathcal{L}_{SFT} = -\frac{1}{|\mathcal{Y}_{pos}|}\sum_{y \in \mathcal{Y}_{pos}} \log \pi_{\theta}(y|x)
\end{equation}

\paragraph{Global Preference Contrast.} We select the top-tier positive sample $y_{w*}$ (e.g., the ordered item) from $\mathcal{Y}_{pos}$ and contrast it with the entire negative pool $\mathcal{Y}_{neg}$ simultaneously. Inspired by softmax-based preference learning \cite{chen2024softmax, guo2025onesug, chen2025onesearch}, we employ a log-sum-exp operator to provide a list-wise contrastive signal. Notably, our formulation is \textbf{reference-free}: by utilizing length-normalized log-probabilities as implicit rewards $\hat{r}_\theta(x,y)$ following SimPO \cite{meng2024simpo}, we eliminate the need for a frozen reference model. This reduces GPU memory footprint by nearly 50\% and streamlines the training process. The preference loss is defined as:


\begin{equation}
\mathcal{L}_{PL}(\theta) = -\log \sigma \left( \beta \cdot \log \sum_{y_j \in \mathcal{Y}_{neg}} \exp(\hat{r}_\theta(x,y_w*)-\hat{r}_\theta(x,y_j)) \right)
\end{equation}
where $\hat{r}_\theta(x,y)$ is defined as:
\begin{equation*}
\hat{r}_\theta(x,y)=\frac{\log\pi_\theta(y|x)}{\vert y \vert}.
\end{equation*}

For efficient implementation, we concatenate the prompt with all candidates into a unified sequence and apply a custom block-diagonal attention mask (as illustrated in Figure \ref{fig-mask}). Crucially, we \textbf{align the position IDs across all candidates} to ensure they share the same relative positional encoding, thereby preventing potential information leakage where the model might distinguish samples based on their relative positions in the concatenated sequence. By rendering the prompt visible to all targets but keeping positive and negative candidates mutually invisible, this structure seamlessly supports session-level multi-label learning. Ultimately, it eliminates computational redundancy by processing all conditionally independent candidates in exactly one forward pass.

\begin{table*}[t]
\centering
\caption{Offline performance comparison on the test set. All DPO-based methods are trained on a subset of 30M search sessions for a fair comparison. Item-level metrics are evaluated after mapping generated SIDs back to corresponding SKUs. Halluc. denotes Hallucination Rate (lower is better).}
\label{tab:main_results}
\begin{tabular}{l|c|cccc|cccc}
\toprule
& \textbf{Halluc.} $\downarrow$ & \multicolumn{4}{c|}{\textbf{Item-level Metrics} $\uparrow$} & \multicolumn{4}{c}{\textbf{SID-level Metrics} $\uparrow$} \\
\cmidrule(lr){3-6} \cmidrule(lr){7-10}
\textbf{Method} & (Rate) & \textbf{R@10} & \textbf{R@100} & \textbf{R@500} & \textbf{MRR} & \textbf{R@8} & \textbf{R@64} & \textbf{R@128} & \textbf{MRR} \\
\midrule
SFT Baseline & \textbf{0.0575} & 0.2958 & 0.5433 & 0.6670 & 0.2415 & 0.3755 & 0.6139 & 0.6775 & 0.2859 \\
DPO & 0.0838 & 0.2731 & 0.5222 & 0.6524 & 0.2286 & 0.3648 & 0.6001 & 0.6642 & 0.2747 \\
SimPO & 0.0708 & 0.3015 & 0.5550 & 0.6747 & 0.2463 & 0.3839 & 0.6214 & 0.6824 & 0.2930 \\
ORPO & 0.0685 & \underline{0.3024} & \underline{0.5579} & \underline{0.6750} & \underline{0.2479} & \underline{0.3844} & \underline{0.6231} & \underline{0.6825} & \underline{0.2940} \\
\midrule
\textbf{RAD-DPO} & \underline{0.0652} & \textbf{0.3048} & \textbf{0.5593} & \textbf{0.6794} & \textbf{0.2481} & \textbf{0.3888} & \textbf{0.6247} & \textbf{0.6864} & \textbf{0.2947} \\
\midrule
\multicolumn{10}{l}{\textit{Ablation Studies}} \\
\quad w/o TLGD & 0.0647 & 0.3025 & 0.5586 & 0.6783 & 0.2470 & 0.3872 & 0.6237 & 0.6864 & 0.2939 \\
\quad w/o RDRW & 0.0657 & 0.3040 & 0.5570 & 0.6792 & 0.2463 & 0.3881 & 0.6226 & 0.6859 & 0.2935 \\
\quad w/o Multi-label SFT & 0.0654 & 0.3031 & 0.5572 & 0.6773 & 0.2455 & 0.3848 & 0.6221 & 0.6849 & 0.2928 \\
\bottomrule
\end{tabular}
\end{table*}

\subsection{Token-Level Gradient Detachment (TLGD)}
In the generation of Semantic IDs (SIDs), positive and negative samples often exhibit strong hierarchical commonalities. For instance, items belonging to the same broad category but different sub-categories share the first $k$ tokens (representing the category path). Standard DPO suppresses the generated probability of the entire negative sequence $y_l$. When $y_w$ and $y_l$ share a common prefix $y_{prefix}$, the gradient $\nabla_{\theta} \log \pi_{\theta}(y_{prefix}|x)$ receives simultaneous push and pull signals. This conflict leads to oscillation during the generation of correct category paths and may damage the hierarchical semantic representations acquired during pre-training.

To resolve this conflict, we introduce a \textit{Prefix Masking} mechanism. This mechanism dynamically truncates the gradient flow of redundant prefixes in negative samples, forcing the model to focus on the critical divergence points between positive and negative samples.
We define an indicator function $1_{diff}(t)$:
\begin{equation}
1_{diff}(t) = 
\begin{cases} 
0, & \text{if } t \le k \ (\text{where } y_{w}^{1:k} = y_{l}^{1:k}) \\
1, & \text{otherwise}
\end{cases}
\end{equation}
where $k$ represents the length of the longest common prefix between the positive and negative samples.
We modify the calculation of the log-likelihood for the negative sample, denoted as $\log \hat{\pi}_{\theta}(y_l|x)$, by applying a gradient detachment operation:

\begin{equation}
\begin{split}
\log \hat{\pi}_{\theta}(y_l|x) = \sum_{t=1}^{L} \bigg( & SG \big[ (1 - 1_{diff}(t)) \log \pi_{\theta}(y_l^t | y_l^{<t}, x) \big] \\
& + 1_{diff}(t) \log \pi_{\theta}(y_l^t | y_l^{<t}, x) \bigg)
\end{split}
\end{equation}
where $SG[\cdot]$ denotes the Stop-Gradient operation.
This operation preserves the forward likelihood value while preventing gradients from back-propagating through the shared prefix tokens of negative samples.

\subsection{Representation-based Dynamic Reward Weighting (RDRW)}

We address the issue of pseudo-negatives, a prevalent source of noise in industrial datasets, by leveraging semantic similarity derived from the model's final hidden states.
Let $h(y, x)$ be the final hidden layer vector when the model generates the End-of-Sequence (EOS) token. For each pair $(y_w, y_l)$, we compute their cosine similarity:
\begin{equation}
sim(y_w, y_l) = \frac{h(y_w, x) \cdot h(y_l, x)}{\|h(y_w, x)\| \|h(y_l, x)\|}
\end{equation}

To ensure the weight adjustment has global awareness, we introduce a statistical warm-up phase. Initially, we cache similarity scores for the first $N=4096$ pairs to construct a baseline distribution $\mathcal{S}$. Based on $\mathcal{S}$, we calculate three key quartile anchors: the first quartile $Q_{25}$, the median $Q_{50}$, and the third quartile $Q_{75}$. While these anchors initially reflect the confusion level between positive and negative samples in the current model state, they are dynamically updated as training progresses to accurately capture the model's shifting semantic representations.

We design a piecewise smooth function to calculate the penalty weight $w(sim)$ for negative samples. This function retains full penalty for distinct negatives while softening the penalty for highly ambiguous samples (potential pseudo-negatives):
\begin{equation}
w(sim) = 
\begin{cases} 
1.0, & sim < Q_{25} \\
0.5 + \frac{0.5}{1 + \exp(\lambda \cdot (sim - Q_{50}))}, & Q_{25} \le sim \le Q_{75} \\
0.5, & sim > Q_{75}
\end{cases}
\end{equation}
where $\lambda$ is set to 12 to control the smoothness of the transition curve.

Finally, the modified preference loss function is expressed as:




\begin{equation}
\begin{split}
\mathcal{L}_{PL}(\theta) &= -\log \sigma \Bigg( \beta \cdot \log \sum_{y_j \in \mathcal{Y}_{neg}} \\
&\quad \exp \big( w(sim)(\hat{r}_\theta(x, y_{w*}) - \hat{r}_\theta(x, y_j))\big) \Bigg)
\end{split}
\end{equation}
where $\hat{r}_\theta$ is defined as:
\begin{equation*}
\hat{r}_\theta(x,y_{w*}) = \frac{\log\pi_\theta(y_{w*}|x)}{\lvert y_{w*} \rvert}, \quad \hat{r}_\theta(x,y_j) = \frac{\log\hat{\pi}_\theta(y_j|x)}{\lvert y_j \rvert}
\end{equation*}
\section{Experiments}
\label{sec:experiments}

\subsection{Experimental Setup}


\textbf{Datasets.} 
We utilize an industrial dataset from JD.com comprising ~700M interaction logs over 8 days, partitioned into 7 days for training and the 8th day $(T+1)$ for testing. Input features encompass the query, user profiles, and historical click sequences. To capture query-relevant interests and mitigate noise, we employ the Search-based Interest Model (SIM \cite{pi2020search}) paradigm with a "hard search" strategy to filter irrelevant items from the click history. The full 7-day dataset is utilized for Supervised Fine-Tuning (SFT), from which a high-quality subset of 30M search sessions is sampled for preference alignment. Crucially, while RAD-DPO models each sample as a complete multi-label search session (e.g., containing both clicks and orders) to optimize global contrast, we decompose these sessions into distinct positive-negative pairs for the baseline DPO methods, ensuring a fair comparative evaluation of partial-order preference learning.

\textbf{Implementation Details.} 
To construct SIDs, we filter the top 170M items with clicks over the past two months. These items are embedded via BGE \cite{bge_embedding} and aggregated into 130M unique SIDs using multi-resolution RQ-Kmeans \cite{he2025plum} with 3-level codebooks (sizes: 8192, 4096, 2048). We employ Qwen-1.7B as the backbone. Notably, all input features---including historical behavior sequences---are injected via structured prompts as SID tokens. 
Training proceeds in two stages: \textit{Supervised Fine-Tuning (SFT)} on click/order paths, followed by \textit{Preference Alignment} via RAD-DPO. Models are optimized on NVIDIA H100 GPUs with a learning rate of $1e^{-6}$ and a batch size of 64.

\textbf{Evaluation Metrics.} Offline ranking accuracy is evaluated using Recall@K and Mean Reciprocal Rank (MRR). Online performance is evaluated via large-scale A/B testing on millions of live users, focusing on User Conversion Rate (UCVR) as the key indicator.

\begin{table}[t]
\centering
\caption{Performance comparison of DPO methods on different base model sizes (SID-level metrics).}
\label{tab:dpo_scaling}

\begin{tabular}{lccccc}
\toprule
\textbf{Model} & \textbf{Method} & \textbf{R@8} & \textbf{R@64} & \textbf{R@128} & \textbf{MRR} \\
\midrule
0.6B & SFT     & 0.3482 & 0.5783 & 0.6343 & 0.2672 \\
     & DPO     & 0.3466 & 0.5732 & 0.6288 & 0.2673 \\
     & RAD-DPO & \textbf{0.3658} & \textbf{0.5867} & \textbf{0.6425} & \textbf{0.2796} \\
\midrule
1.7B & SFT     & 0.3755 & 0.6139 & 0.6775 & 0.2859 \\
     & DPO     & 0.3589 & 0.6001 & 0.6642 & 0.2747 \\
     & RAD-DPO & \textbf{0.3888} & \textbf{0.6247} & \textbf{0.6864} & \textbf{0.2947} \\
\midrule
4B   & SFT     & 0.3992 & 0.6600 & 0.7235 & 0.3080 \\
     & DPO     & 0.3880 & 0.6521 & 0.7159 & 0.2983 \\
     & RAD-DPO & \textbf{0.4178} & \textbf{0.6722} & \textbf{0.7341} & \textbf{0.3161} \\
\midrule
8B   & SFT     & 0.4327 & 0.6924 & 0.7436 & 0.3178 \\
     & DPO     & 0.4128 & 0.6837 & 0.7381 & 0.3152 \\
     & RAD-DPO & \textbf{0.4611} & \textbf{0.6983} & \textbf{0.7492} & \textbf{0.3246} \\
\bottomrule
\end{tabular}
\vspace{1em}
\end{table}

\subsection{Results and Discussion}

\begin{figure}[t]
    \centering
   \includegraphics[width=\linewidth]{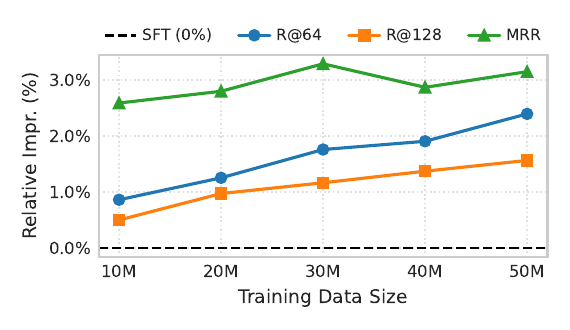}

    \caption{Impact of DPO training data scale on model performance. Results are presented as relative percentage improvements over the SFT baseline across key SID-level metrics, including Recall@K and Mean Reciprocal Rank (MRR).}
    \label{fig-datascale}
    \vspace{2em}
\end{figure}

\textbf{Overall Performance.} As detailed in Table \ref{tab:main_results}, RAD-DPO achieves consistent improvements in both Recall and MRR over the SFT baseline and standard DPO, confirming that mitigating the unintended suppression of positive signals is essential for generative retrieval. Beyond these top-line metrics, the framework exhibits remarkable scalability and data efficiency. Not only does its performance edge widen as model capacity increases from 0.6B to 8B parameters (Table \ref{tab:dpo_scaling}), but it also sustains robust ranking quality even across reduced training scales (10M to 50M, Figure \ref{fig-datascale}). 
Finally, RAD-DPO also lowers DPO’s hallucination rate by retaining token-level constraints and the SFT loss.

\textbf{Ablation Study.} As detailed in Table 1, removing any core component of RAD-DPO leads to noticeable performance degradation, validating their distinct contributions. Specifically, discarding Token-Level Gradient Detachment (\textit{w/o TLGD}) drops the MRR, confirming its effectiveness in resolving training oscillations caused by shared hierarchical prefixes. Furthermore, the performance decline without Dynamic Reward Weighting (\textit{w/o RDRW}) underscores the disruptive impact of pseudo-negatives in raw industrial logs—a noise successfully neutralized by our adaptive penalty softening. Finally, removing the Multi-label Global Contrast (\textit{w/o MLGC}) distinctly reduces Recall, demonstrating that replacing myopic pairwise comparisons with global multi-positive signals is essential for maximizing preference learning throughput.


\textbf{Online A/B Testing.} To evaluate the real-world impact, we conducted online A/B testing on the \textbf{core e-commerce search engine} of JD.com, which serves hundreds of millions of daily users. RAD-DPO was implemented as a standalone generative retrieval branch that bypasses traditional multi-stage retrieval and pre-ranking. Specifically, it directly generates item candidates that are then merged with outputs from other branches at the final ranking stage. Crucially, the 1.7B model meets strict industrial serving requirements, achieving a prediction latency of $<150$ms (beam width=128) at 15 QPS on a single NVIDIA RTX 4090 GPU. This week-long functional ablation test demonstrated a consistent \textbf{+0.34\%} lift in User Conversion Rate (UCVR). Following this success, the model was re-trained on an extended dataset of hundreds of millions of logs and rebased with other production features for a formal full-scale launch.

\section{Conclusion}


In this work, we proposed RAD-DPO to address the systematic suppression of positive semantic signals when applying standard DPO to Generative Retrieval in e-commerce search. By introducing multi-label global contrast, token-level gradient detachment, and dynamic reward weighting, our framework effectively resolves gradient conflicts and neutralizes pseudo-negative noise. Extensive evaluations on a large-scale industrial e-commerce platform demonstrate that RAD-DPO significantly enhances retrieval precision, while the design of a custom block-diagonal attention mask substantially boosts training efficiency. More importantly, online A/B testing confirms that it delivers measurable business conversion lifts while strictly meeting industrial serving latency requirements, offering a robust paradigm for structured preference alignment.

\clearpage

\printbibliography

\section*{Company Portrait}
JD.com, Inc., also known as Jingdong, is a Chinese e-commerce company headquartered in Beijing. It is one of the two massive B2C online retailers in China by transaction volume and revenue, a member of the Fortune Global 500. When classified as a tech company, it is the largest in China by revenue and 7th in the world in 2021.

\section*{Presenter profiles}
\noindent\textbf{Zhiguo Chen} is a researcher in the Department of Search and Recommendation at JD.com Beijing. 
His research focuses on information retrieval, cross-modal alignment and natural language processing.

\noindent\textbf{Guohao Sun} is a Master's student at Peking University. His research interests lie at the intersection of Large Language Models and search-related technologies, with a secondary focus on reinforcement learning. Currently, he is a research intern at JD.com.

\end{document}